\documentclass{article}[10pt]
\usepackage[utf8]{inputenc}
\usepackage{url}
\usepackage{booktabs}
\usepackage{graphicx}
\usepackage{tabularx}
\usepackage{ltablex}
\usepackage{tabu}
\usepackage[affil-it]{authblk}

\usepackage{tabulary}

\usepackage[newcommands]{ragged2e}
\usepackage{blindtext}
\usepackage{geometry}
\usepackage{titlesec}

\titleformat{\paragraph}
{\normalfont\normalsize\bfseries}{\theparagraph}{1em}{}
\titlespacing*{\paragraph}
{0pt}{3.25ex plus 1ex minus .2ex}{1.5ex plus .2ex}

\begin{document}
 
\title{BactInt: A domain driven transfer learning approach and a corpus for extracting inter-bacterial interactions from biomedical text}


\author[1, 2]{Krishanu Das Baksi}
\author[1]{Vatsala Pokhrel}
\author[1, *]{Kuntal Kumar Bhusan}
\author[1]{Sharmila Mande}

\affil[1]{TCS Research, Pune, India}
\affil[2]{School of Information Technology, IIT Delhi, India}
\affil[*]{Corresponding author; email: kuntal.bhusan@tcs.com}

\date{April 2023}
\maketitle

\begin{abstract}
The community of different types of microbes present in a biological niche plays a very important role in functioning of the system. The crosstalk or interactions among the different microbes contributes to the building blocks of such microbial community structures. Evidence reported in biomedical text serves as a reliable source for predicting such interactions. However, going through the vast and ever-increasing volume of biomedical literature is an intimidating and time consuming process. This necessitates development of automated methods capable of accurately extracting bacterial relations reported in biomedical literature. In this paper, we introduce a method for automated extraction of microbial interactions (specifically between bacteria)  from biomedical literature along with ways of using transfer learning to improve its accuracy. We also describe a pipeline using which relations among specific bacteria groups can be mined. Additionally, we introduce the first publicly available dataset which can be used to develop bacterial interaction extraction methods.

\end{abstract}

\section{Introduction}
Microbes are one of the most important groups of organisms in all environments with bacteria being the dominant part. Understanding the microbial community structure and the roles that they play in their respective environments is a topic of interest in a diverse number of fields, including but not limited to medicine, agriculture, nutrition etc \cite{microbeplant, microbehealth, microbehealth3}. Such understanding can help us decipher more about an environment, and also potentially reveal strategies to manipulate the same for a desired objective. In recent years, with the realization of the importance of the microbiome, scientists have carried out a large number of experiments to obtain data pertaining to bacterial composition in different ecosystems \cite{earthmicrobe}. However, relationships inferred from such data using techniques like correlation between the count abundances are often subjected to certain limitations owing to the compositional nature of the data \cite{compositional} as not all correlations can be ascertained to be interactions. Experimentally reported microbial interactions in biomedical literature still serve as the gold standard. Extraction of such information from biomedical text can provide deeper insights into the microbial communities and potentially suggest how to manipulate them for industrial \cite{microbeindustry} and healthcare benefits \cite{microbehealth, microbehealth3}. However, due to the exponential increase in the size and scale of scientific literature \cite{pubmedgrowth}, mining of this information manually is time consuming. Advanced search engines can partially help in this regard by finding the documents or in some cases, sentences containing mentions of one or more microbial names. However, the final step of inferring inter bacterial associations and interactions can only be achieved using specialized techniques that work on top of the search engines. Methods have been developed for the identification of abstracts and sentences reporting microbial associations using hand-crafted domain features \cite{bacmass}. Such methods can identify only the sentences reporting microbial relations without extracting the actual relations among them. Realizing the importance of extraction of inter-microbial associations from text, several scientific works has been done for mining the same \cite{evimass, mplasso}. For example, EviMass \cite{evimass} and MPLasso \cite{mplasso} identify significant associations among microbes using their co-occurrence statistics, i.e., how many times they are mentioned in the same biomedical abstract. However, instead of extracting the actual reported interactions, these methods rely on the co-occurence statistics of microbial names in scientific literature. Although the co-occurence statistics may indicate some form of association (direct or indirect) between microbes, it is difficult to predict whether these associations imply any direct biological interactions. 

\begin{figure*}[t]
    \label{fig:fig1}
    \centering
    \fbox{\includegraphics[width=0.9\textwidth]{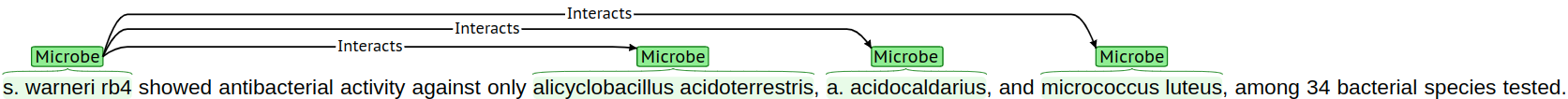}}
    \caption{Example of an intermicrobial-interaction annotation, with marked bacterial entities and the interacting pair.}
\end{figure*}

Recent advances in natural language processing, especially the field of information extraction holds promise for identifying microbial relations from scientific literature using automated techniques. Information extraction broadly refers to the methods for extracting relationships among different entities from natural language text. Current deep learning based information extraction approaches, albeit highly accurate, require large amount of manually labelled training data to perform well. However, creation of a manually annotated dataset is a time and effort consuming process. Recent methods aimed at extracting inter-bacterial interactions or relations from biomedical scientific literature text has also been described \cite{r13, r14}. However, the datasets used to train the models are not available publicly.

In this paper, we present a manually annotated dataset which can be utilized for developing machine learning methods for extracting bacterial interaction, and make it available publicly. Since the datasets are manually curated by highly trained domain experts, collecting a very large dataset was prohibitively expensive and time consuming. Consequently, the dataset we developed is limited in size, containing around 1400 data points. Usually, the performance of modern NLP models based on deep learning suffer when the datasets are not very large. In order to ensure high performance of the information extraction method, we devise a transfer learning strategy which exploits other publicly available biomedical information extraction datasets, and show its effectiveness. Finally, we also present an end-to-end pipeline, including sentence segmentation, named entity recognition and information extraction for finding bacterial interactions reported in biomedical text.

\subsection{Contributions}

In this paper, we make the following contributions: 

1. We present a manually annotated dataset for extraction of bacterial interactions from sentences belonging to relevant biomedical scientific literature, and make it publicly available. To the best of our knowledge, this is the first publicly available dataset pertaining to extraction of inter bacterial interactions from biomedical text. 

2. We also develop and train a BERT \cite{bert} based information extraction model for the task. We acknowledge that the size of the dataset is not large, and that it is a limiting factor in the performance of the model, and additionally demonstrate an \textit{explicit transfer learning} strategy for improving the model’s performance significantly, exploiting other publicly available biomedical relation extraction datasets.

3. We present an end-to-end pipeline for finding the interactions or relations among bacterial taxa in a biomedical text (e.g., biomedical literature abstracts or paragraphs), and evaluate the performance of each of the components of the pipeline.

We expect that our contribution will serve as a useful methodology for microbiome researchers and engineers to find the interactions among microbes of their interest. Additionally, we hope that the models and annotated data presented in this paper will serve as the base for the development of better techniques and architectures for the extraction of microbial interactions from biomedical text. Interested readers are requested to get in touch with the corresponding author for accessing the data.

\begin{figure*}[t]
    \label{fig:diff_transfer_learning}
    \centering
    \fbox{\includegraphics[scale=0.6]{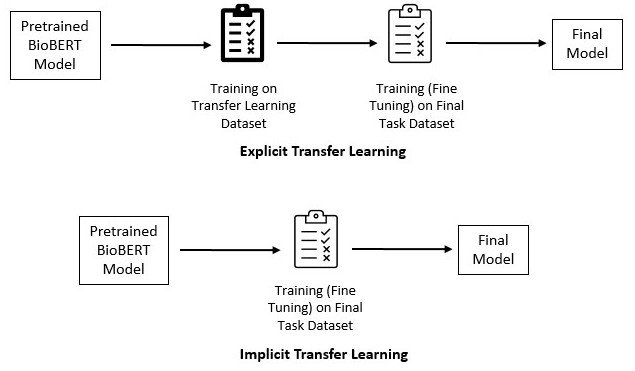}}
    \caption{Difference between implicit and explicit transfer learning training regimens as followed in this work.}
\end{figure*}

\section{Background}

\subsection{Information Extraction in Biomedical Domain}
Knowledge in the domain of biological sciences is primarily available in unstructured text form, making it difficult for computer systems to extract useful information from the same \cite{r1, r2}. To enable mining and inference of new knowledge using known information present in these documents, it is necessary to capture this data and make it available in a structured format \cite{r2}.

Several studies and datasets were introduced in order to solve this problem in multiple subdomains of biomedical sciences. Since protein-protein interactions are very useful for many applications, multiple methods for extraction of the same from from biomedical literature have been reported \cite{r3, r4, r5, r6}. Apart from protein protein interactions, datasets for the extraction of drug-drug interactions, viz ADE \cite{r7} and DDI \cite{r8}, chemical-protein interactions viz. ChemProt \cite{r9} and several others are also available.

\subsection{Methods for Information Extraction}

Different types of methods have been developed for information extraction in the domain of biology. In the past, most of these methods relied on hand-crafted features \cite{r27, r18, r20} and rules \cite{r15, r17, r19, r21}. Creation of hand-crafted rules or features is a tedious process as a large number of possible syntactic (grammatical) variations \cite{r16} need to be taken into account. Therefore, with the rise in popularity of deep learning and the creation of large scale datasets, the focus shifted towards deep learning based methods which are capable of automatically learning features \cite{r22}. Currently, transformer \cite{transformer} based architectures like BERT \cite{bert} outperform older neural architectures like recurrent neural networks and tree based recursive neural networks on several NLP problems including bacterial relation extraction \cite{r14}.

\subsection{BERT and BioBERT} BERT \cite{bert, biobert} makes use of encoder block of the transformer \cite{transformer} architecture. It is based on the self-attention mechanism, which learns contextual relations between tokens (words or subwords) in a text. The BERT architecture and its constituent layers, as well as methods for unsupervised pretraining the model is described in detail in the original papers on transformers \cite{transformer} and BERT \cite{bert}. The pretrained BERT models usually show strong performance \cite{bert} on multiple NLP tasks. Different models pertaining to multiple domains and languages which are pretrained using the BERT methodologies have been described in literature. In the biomedical domain, BERT model pretrained on biomedical domain specific scientific literature called BioBERT \cite{biobert} has been described, and has been used for a variety of downstream applications in the domain of biomedical NLP. 

\subsection{GPT and BioGPT} GPT \cite{gpt, gpt2, biogpt} makes use of the decoder block of the transformer \cite{transformer} architecture and is trained as an generative language model. Whereas the BERT model by design can primarily find contextual embeddings of the tokens in an input text, GPT like models have the additional capacity of generating text. Additional details into the model architecture, training methodology and other details can be found in the original papers \cite{gpt, gpt2}. Like BERT, pretrained GPT models have also been shown to perform very well on a number of NLP, especially generative NLP tasks. The popular ChatGPT system is also based on the GPT architecture. Recently, BioGPT \cite{biogpt}, a GPT2 \cite{gpt2} based model pretrained on biomedical scientific literature text has been shown to perform well on a variety of biomedical NLP tasks.

\subsection{Transfer Learning}

Transfer learning pertains to training (learning the parameters of) a model on one task and then relearning or retraining the model on another related (having some common underlying structure) but different task. Models show improved performance on the final task after transfer learning, and can also work well when limited training data is available for the final task \cite{transferlearning}. The underlying rationale behind this pertains to the fact that the model can utilize what it 'learned' in the first task to perform better on the second task. Here it is important to note that for most BERT and GPT based neural architectures, transfer learning is \textit{implicit}, as the 'learning' during the pretraining step is 'transferred' while the model is being fine-tuned on the final task. For example, the effectiveness of a BERT model pretrained using biomedical literature text (such as BioBERT \cite{biobert} or SciBERT \cite{scibert}) and further trained / fine-tuned on the task of extraction of microbial interactions from text has been demonstrated \cite{r14}. This type of transfer learning is \textit{implicit} in the pretraining methodology. However, transfer learning can also be more \textit{explicit}, for example, a BERT model already pretrained using biomedical literature text (such as BioBERT \cite{biobert}) can be further trained for the prediction of protein-protein interactions from text and further on the task of microbial interaction extraction from text. In the context of the current work, this distinction between \textit{explicit} and \textit{implicit} transfer learning is very important and worth noting. It has also been explained using Figure \ref{fig:diff_transfer_learning}.

\textbf{Implicit transfer learning} refers to the methodology where the pretrained model is fine tuned on any downstream task. In this strategy, the model uses what it learnt during the initial pretraining process, for example, during the masked language modelling and next sentence prediction tasks in case of BERT, on the final downstream task. Fine-tuning a BioBERT \cite{biobert} or BioGPT \cite{biogpt} model on any downstream biomedical NLP task would belong to this category.

\textbf{Explicit transfer learning} refers to explicitly training the already pretrained models on a task which is somewhat similar in structure to the final task in certain characteristics. An example would be training a (already pretrained) BioBERT or BioGPT model for the task of protein-protein interaction extraction task, and then fine-tuning on microbe-microbe interaction task. This example methodology is also the primary strategy followed in this work, in order to improve performance of the models on limited annotated data. In comparison to the \textbf{implicit transfer learning} method, there is an additional training regimen, but which is focused on a task similar (in structure) to the one at hand, for which substantial amount of data is available.

\subsection{Information Extraction related to Bacteria and Microbes}

Since the relations among microbes can provide insights into the complex interactions among individual strains and species in a microbiome, datasets have also been developed for the extraction of microbial relations from biomedical text. In \cite{r13} and \cite{r14} two corpora (viz. MICorpus and MTMICorpus) focusing on microbe interaction extraction along with deep learning methods for the same have been described. While the MICorpus only focuses on the task of extraction of interacting microbes, MTMICorpus also contains the type of interaction between a pair of interacting microbes \cite{r14}. The performance of microbial relation extraction using implicit transfer learning, wherein a BioBERT \cite{biobert} and SciBERT \cite{scibert} models (pretrained on biomedical scientific literature) are fine tuned for the dataset specific for the task has been illustrated \cite{r14}. However, it is important to note that the datasets described in these papers are not available publicly, therefore hindering further experimentation and progress towards solving this problem.

\begin{figure*}[!htp]
    \label{fig:fig2}
    \centering
    \fbox{\includegraphics[width=0.9\textwidth]{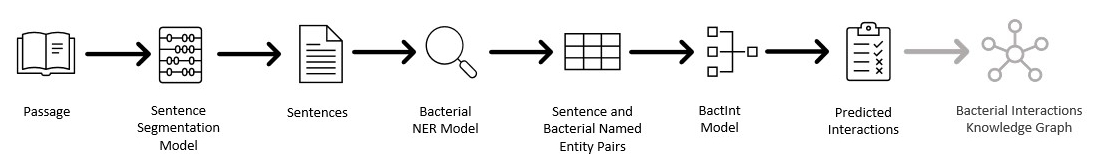}}
    \caption{The End-to-End Pipeline for bacterial interaction extraction from a given biomedical text passage. 
    \\      
    N.B. The last step is not done in this paper.}
\end{figure*}

\section{Methods}

\subsection{Bacterial Interaction Extraction Sentence Corpus}
The bacterial interaction sentence corpus, which we refer to as the 'BactInt Sentence Corpus' is composed of around 1400 sentences annotated with all bacterial mentions as well as their interactions, if any. The primary motivation for annotating individual sentences rather than entire abstracts is that only a few sentences per abstract indicate some kind of interaction. Since annotation is done by highly trained biomedical experts, this is expensive, and annotating entire abstracts would essentially need annotating multiple sentences with no reported interaction, and probably a few with a reported interaction. This would take up a lot of time, but the number and diversity of sentences which report an annotation would not be optimized. However, in order to make the model generalize well, the goal to maximize the number and diversity of our dataset. Therefore, only sentences were annotated. The process is detailed in the following sections. 

\subsubsection{Identification of Sentences for Annotation}
The dataset of sentences reported in \cite{bacmass} consisting of around 750 sentences, wherein multiple bacterial names are mentioned in each sentence, is taken as the initial dataset. Out of these sentences, while half of them report at least one bacterial interaction (some report multiple) between the mentioned bacterial entities, the other half do not report any bacterial interaction. In addition, we identify several other sentences using the following method. A set of abstracts from PubMed (\url{https://pubmed.ncbi.nlm.nih.gov}) containing multiple bacterial mentions is selected, and subset of these abstracts are further filtered using an 'Abstract classifier' described in \cite{bacmass}. These abstract passages are broken down into individual sentences using SpaCy (\url{https://spacy.io}), specifically the 'en\_core\_sci\_md' (SciSpacy \cite{scispacy}) model. Finally, the sentences having a high probability of reporting a bacterial interaction are selected from this set using another 'Sentence classifier' described in \cite{bacmass}. A total of around 650 sentences are identified. The combined set of these sentences with the initial ones (total 1406 sentences) are further annotated.

\subsubsection{Manual Annotation of the Sentences}

Each of the sentences were annotated manually by biomedical experts. The well-known open source information extraction annotation tool BRAT \cite{brat} was used for this purpose. The experts annotated every mention of bacterial names in the text, by labelling the span of the text that corresponded to a bacterial named entity. To infer whether the bacterial entities present in a sentence have an interaction, the sentences were required to have a clear mention of two or more bacterial entities exhibiting an effect on each other, which were then finally annotated as an inter-bacterial interaction. Examples of the nature of such inter-bacterial interactions include bacterial entities inhibiting the growth of the other bacterial entities, bacteria promoting the growth of the other bacterial entities, bacterial entities having a synergistic or antagonistic effect on each other, etc., all of which are collectively labelled as ‘interacts’. For instance, if a sentence has clear mention of two bacterial entities, ‘bac1’ and ‘bac2’, and ‘bac1’ has any kind of effect (inhibitory, growth promoting, synergistic, antagonistic, etc.) on ‘bac2’, the annotation was labelled as ‘bac1’ ‘interacts’ with ‘bac2’, indicating an inter-bacterial interaction for that particular sentence. It must be noted that each text (sentence) can potentially have zero, one or multiple associations. In order to detect the maximum number of potential bacterial associations from biomedical text, the sentence corpus captures information from a diverse range of environment ranging from human host, marine ecosystem, soil microbiome, plant symbionts, etc. \cite{bacmass}. Fetching information from such diverse environments enables us to understand the various mechanisms of bacterial interactions that the bacterial entities utilize for survival like biofilm formation, quorum sensing, production of volatile organic compounds, bacteriocins, polyketides, toxins, siderophores, etc. Thus, the manually annotated corpus had sentences capturing information from diverse environments involving bacterial interactions with the above-mentioned mechanisms of associations. It is important to note that out of the various inter-microbial associations present in different environments, only bacteria-bacteria associations have been considered for the current work and information pertaining to the viral, fungal and protozoan interactions have not been included. A glimpse (visualization) of an annotated sentence can be found in Figure \ref{fig:fig2}. This dataset is henceforth referred to as the 'BactInt Sentence Corpus'.

\subsection{Bacterial Interaction Extraction End to End Corpus}
The BactInt Sentence corpus is designed to train models which can predict the interactions reported among bacterial named entities in a single sentence. However, an end-to-end pipeline, which can take in a text passage, for e.g. a scientific abstract and predict all the individual interactions reported in the entire abstract, is desirable. Such an end-to-end pipeline would be typically composed of a sentence tokenizer, which breaks down the passage into a sentence, followed by a bacterial named entity recognition methodology which tags the bacterial mentions in the sentences, followed by a methodology that can identify the interactions reported among a pair of bacterial mentions in a given sentence.

Since the end-to-end pipeline is composed of several individual components with each having its set of limitations, the performance of the entire pipeline also need to be evaluated. For this, a small set of 25 biomedical scientific literature abstracts containing multiple bacterial named entity mentions were selected. Among these abstracts, 10 report at-least one microbial interaction and 15 report no microbial interactions. The bacterial named entities and the interactions reported among them were manually annotated. Care was taken to ensure that each of the annotated interactions can be inferred from one particular sentence of the passage. This was done to ensure that the dataset does not contain interactions that need multiple sentences together to be inferred, as this is beyond the scope of the current methodology. This dataset is henceforth referred to as the 'BactInt E2E Corpus'. This dataset is quite small compared to the sentence corpus, as its utility is not to train any models but to understand the performance of the pipeline, and analyse the possible sources of errors.


\subsection{Transfer Learning Datasets}
In order to exploit available biomedical information extraction data with the goal of improving the performance of the model, the following corpora were identified: BioInfer \cite{bioinfer}, HPDR50 \cite{hpdr50}, LLL \cite{lll}, IEPA \cite{iepa}, AiMED \cite{aimed}, Gene Regulation \cite{genereg} and DDI 2011 \cite{ddi2011}. While BioInfer, HPDR50, LLL, IEPA and AiMED pertain to protein protein interactions, the Gene Regulation corpus, as the name suggests, is targeted towards extraction of gene regulatory information, and DDI 2011 corpus focuses on extraction of reported drug-drug interactions. Some of the text passages provided in the datasets were composed of multiple sentences. However, the BactInt Sentence Corpus focuses on extraction of interactions from a single sentence at a time. In order to make the pretraining corpus similar to the BactInt Sentence Corpus, each individual text in the corpus were broken down into individual sentences by SpaCy (\url{https://spacy.io}), specifically the 'en\_core\_sci\_md' (SciSpacy \cite{scispacy}) model. All the annotations (named entities and interactions/relations) present in a single sentence were added. Annotations reported interactions/relation among entities present in different sentences were ignored. In total there were more than 9 thousand individual sentences reporting more than 8 thousand interactions (each sentence may have zero, one or multiple reported interactions/relations). This combined dataset is henceforth referred to as the 'Pretraining Corpus'.

\subsection{Our Approach - Explicit Transfer Learning}

\subsubsection{Model}
Since Transformer \cite{transformer} based neural network architectures like BERT \cite{bert} and GPT \cite{gpt} are known to outperform older neural architectures like recurrent neural networks and tree based recursive neural networks on several NLP problems including bacterial relation extraction, in this paper, the performance of BioBERT \cite{biobert}) as well as BioGPT \cite{biogpt} is analysed, and other deep learning architectures are ignored. The BioBERT and BioGPT models were selected as the base model and was further used in \textit{implicit transfer learning} as well as \textit{explicit transfer learning} settings, as explained in the subsequent sections. For all our experiments, we used the implementation (of BERT and BioGPT) and model parameters (of BioBERT and BioGPT) from the HuggingFace library (https://huggingface.co). 

\subsubsection{Methodology}
In our approach, we first trained the BioBERT and BioGPT models on the task of relation extraction between biomedical entities from biomedical text sentences, using the \textit{transfer learning} datasets as explained in a previous section, viz. BioInfer \cite{bioinfer}, HPDR50 \cite{hpdr50}, LLL \cite{lll}, IEPA \cite{iepa}, AiMED \cite{aimed}, Gene Regulation \cite{genereg} and DDI 2011 \cite{ddi2011}. After combined training on this first set of transfer learning datasets, the models were fine tuned on the BactInt Sentence dataset. This training regime is referred to as \textit{explicit transfer learning} and the trained models are referred to as \textit{BioBERT Explicit} and \textit{BioGPT Explicit}.


\subsection{Data Transformation for Training and Inference}

For every text or data point in our annotated corpora, we transformed the data such that the information extraction problem can be converted into a binary classification problem \cite{relex_to_classification}. In brief, each sentence is tagged using the annotated bacterial named entity information. For every pair of unique bacterial named entities, a data-point (sentences with suitably tagged pair of entity mentions) is created, and they are assigned a binary target variable based on whether an interaction is annotated between the aforementioned pair of entities.





The interaction extraction machine learning model had to take the transformed data-point as input and essentially classify it to the correct target category. Therefore, the task of information extraction was converted to binary classification, a well known machine learning problem.

\subsection{Data Split}

The 'BactInt Sentence Corpus' dataset was randomly split into two datasets, the larger comprising of 85\% annotated sentences was used for training, and a smaller set of 15\% annotated sentences was used for testing the performance of the trained models.

\subsection{Training}

The training of the BioBERT and BioGPT model was carried out in the following way:

1. The input data was transformed as explained above, before being input to the BioBERT or BioGPT model.

2. The embedding vector of the "[CLS]" token computed by the BioBERT model was used for classification. On the other hand, for BioGPT model, the embedding of the last token ("<s/>") was used for classification. The embeddings produced by the BioBERT or BioGPT model were passed through logistic regression classifier heads (learn-able linear transformation followed by a sigmoid activation function) to produce the final output score (which can be thought of as the probability of the positive class). 

3. The loss between the predicted output score (probability) and the target was computed using the cross-entropy loss function. 

4. The loss was minimized using back-propagation, like most neural classifier methods. This in turn updates the model parameters of the BioBERT and BioGPT models (along with the parameters of the classifier head). This is done for several passes through the entire training dataset.

5. Early stopping was used to reduce over-fitting.

\subsection{Baselines}
\label{baseline}
\subsubsection{Implicit Transfer Learning Models} In this category, BioBERT and BioGPT models were fine tuned on the final task viz. bacterial interaction extraction, without any explicit transfer learning step. Therefore, the training methodology, in principle, is identical to the one described in \cite{r14}. As explained before, since these models have been pretrained on biomedical literature corpus, it is a transfer learning model, and is referred to as \textit{implicit transfer learning} models. The BioBERT and BioGPT models trained using this methodology are named \textit{BioBERT Implicit} and \textit{BioGPT Implicit}.

\subsubsection{Non Fine-Tuned Models} In order to analyse how much the models learn from the transfer learning tasks, we train both the BioBERT and BioGPT models on the transfer learning datasets, and without any fine tuning, analyse their performance on the bacterial information extraction task. These models are named \textit{BioBERT Non-FT} and \textit{BioGPT Non-FT} respectively.


\subsection{End-to-End Bacterial Interaction Extraction Pipeline}
\label{method_e2e}
The models discussed in the previous sections can only extract reported bacterial interactions in individual sentences where the bacterial named entities are already marked / annotated. However, in practice, end-users would be interested in extracting the interactions reported in a passages (consisting of several sentences without any bacterial named entities). Therefore, the bacterial interaction extraction model from sentences would need some auxiliary methods, viz. a bacterial named entity recognition method, which can identify and mark all the bacterial named entities in the provided text, and a sentence tokenizer, which can accurately break down the entire passage into individual sentences. The model can then take in the individual sentences with marked bacterial named entities as input and predict the interactions reported, if any, as output.

The following methods were used in the end-to-end pipeline. A schematic diagram is provided in Figure \ref{fig:fig2}:

1. The SciSpacy library \cite{scispacy}, specifically the 'en\_core\_sci\_md' model was used for segmenting a passage into a set of individual sentences.

2. A BioBERT \cite{biobert} based microbial named entity recognition model (BioBERT BNER)\cite{bacnerlm}  was used to predict the bacterial named entities.

3. The bacterial interaction extraction model viz. \textit{BioBERT Explicit} or \textit{BioGPT Explicit} was used to extract the interactions from the individual sentences with marked bacterial named entities as described previously.

\subsubsection{Bacterial Named Entity Recognition Model}
The BioBERT based bacterial NER model described in \cite{bacnerlm} was used to extract the bacterial named entities from the individual sentences. Since the trained model and code were unavailable publicly, the dataset described in the paper \cite{bacnerlm} was used to train a BioBERT based bacterial named entity recognition model, based on the descriptions found in the paper \cite{bacnerlm}.



\begin{table*}[]
\label{tab:primary}
\caption{Comparison of the performance of the different training methodologies on the test set. The average value of each of the metrics over 3 runs and the standard deviations are provided.}
\begin{tabular}{|l|l|l|l|l|l|l|}
\hline
          & \begin{tabular}[c]{@{}l@{}}BioBERT \\ Implicit \\ Not Fine Tuned\end{tabular} & \begin{tabular}[c]{@{}l@{}}BioBERT \\ Implicit\end{tabular} & \begin{tabular}[c]{@{}l@{}}BioBERT \\ Explicit\end{tabular} & \begin{tabular}[c]{@{}l@{}}BioGPT \\ Implicit \\ Not Fine Tuned\end{tabular} & \begin{tabular}[c]{@{}l@{}}BioGPT \\ Implicit\end{tabular} & \begin{tabular}[c]{@{}l@{}}BioGPT \\ Explicit\end{tabular} \\ \hline
Precision & 0.7 $\pm$ 0.02                                                                   & 0.84 $\pm$ 0.03                                                & 0.87 $\pm$ 0.03                                                & 0.61 $\pm$ 0.13                                                                 & 0.78 $\pm$ 0.07                                               & 0.83 $\pm$ 0.02                                               \\ \hline
Recall    & 0.7 $\pm$ 0.07                                                                   & 0.77 $\pm$ 0.02                                                & 0.9 $\pm$ 0.02                                                 & 0.56 $\pm$ 0.13                                                                 & 0.8 $\pm$ 0.06                                                & 0.87 $\pm$ 0.05                                               \\ \hline
F1        & 0.7 $\pm$ 0.03                                                                   & 0.8 $\pm$ 0.01                                                 & 0.89 $\pm$ 0.02                                                & 0.57 $\pm$ 0.08                                                                 & 0.79 $\pm$ 0.06                                               & 0.85 $\pm$ 0.02                                               \\ \hline
\end{tabular}
\end{table*}

\section{Experiments and Results}

\subsection{Evaluating the Gain in Performance due to Explicit Transfer Learning}
In order to evaluate the utility of our approach, viz. Explicit Transfer Learning, BioBERT \cite{biobert} and BioGPT \cite{biogpt} models were first trained on the 'Pretraining Corpus', followed by fine-tuning on the task-specific 'BactInt Sentence Corpus' training split. These model trained using this methodology are referred to as 'BioBERT Explicit' and 'BioGPT Explicit'. Additionally baseline models were also trained as explained in Section \ref{baseline}. The precision, recall and F1 score of each of the models on the test dataset were computed. In order to take into account random variations (due to steps like model initialization, training-validation split etc.), each of the training steps were repeated three times. A comparison of the performance of the models can be found in Table \ref{tab:primary}. Overall, it can be observed that the \textit{explicit transfer-learning} models outperform the \textit{implicit transfer learning} and the non fine-tuned models by a significant margin. Results clearly indicate a significant improvement in performance due to the explicit transfer learning methodology. Moreover, it can be seen that the BioBERT explicit model performs better than BioGPT explicit model. Also, it is interesting to note that the non-fine-tuned models also show decent performance. This highlights the positive impact of the transfer learning tasks on the final bacterial interaction extraction task. In case of non fine-tuned models, BioBERT shows better performance as compared to BioGPT, which also explains the reason why the BioBERT explicit model outperforms the BioGPT explicit model.

\subsection{Evaluating the Performance of the End-to-End Pipeline }

The end-to-end pipeline was constructed as described in the Section \ref{method_e2e}. As the bacterial interaction extraction model, the BioBERT Explicit models were used, as they were the best performing models in the previous evaluation. The interactions predicted by the pipeline were compared to the manually annotated ones. Firstly, a pipeline with all components, viz. the sentence segmentation ("SS"), named entity recognition and tagging ("NER"), and information extraction ("IE") models was used, which is referred to as "SS + NER + IE" . In order to evaluate the loss in performance due to the sentence segmentation module, a pipeline which consisted of the NER model and the information extraction model, but without the sentence segmentation module was used, which is referred to as "NER + IE". The gold standard ground truth sentences from the end-to-end corpus were provided to this pipeline. Finally, a pipeline which consisted of only the information extraction model was used, in order to evaluate the loss in performance due to the NER and SS modules. This is referred to as "Only IE", and the ground truth sentences with the ground truth named entity tags were provided to this pipeline. The precision, recall and F1 score of each of the pipelines were recorded and depicted in Table \ref{tab:e2e}. This analysis gives an indication of the loss in performance caused by each of the different components. For e.g. the performance drop caused by the NER module can be inferred by comparing the performance of the "NER + IE" and the "Only IE" pipelines. Similarly, the performance drop caused by the SS module can be inferred by comparing the performance of the "SS + NER + IE" with that of "NER + IE" modules. Finally, the performance drop due to the IE module can be inferred by the performance of the "Only IE" pipeline.

Analysing the results in Table \ref{tab:e2e} reveals that the precision falls significantly (0.89 vs 0.63) when the pipeline uses the NER model ("NER + IE") as compared to the one where ground truth named entities are provided ("Only IE"). Although recall also falls (0.93 vs 0.84) but the drop is not as drastic as that of precision. This indicates that the BioBERT NER model is a major source of errors, and that it produces more false positives than false negatives. A closer manual examination of the results reveals that the NER model often predicts non-bacterial names as bacterial named entities. For e.g. the NER model often predicted terms representing non-bacterial taxa like 'Caenorhabditis elegans' as bacterial named entities. Here it is important to note that the IE model inherently doesn't differentiate between bacterial and non-bacterial entities and primarily aims to predict the interactions between the tagged named entities. Therefore if an interaction exists between a bacterial and a non-bacterial entity, if the NER model predicts the non-bacterial entity as a bacterial one, then the IE model will also predict an interaction, leading to an erroreous prediction. This type of errors cause a large drop in precision of the "IE + NER" pipeline in comparison to the "Only IE" pipeline. In addition to these, it was seen that in very rare cases, the sentence segmentation module segments the sentences incorrectly. For example, the sentence segmentation model at times splits sentences between abbreviated bacterial named entities like 'Lb. oligofermentans' (Lactobacillus oligofermentans), 'Lc. piscium' (Lactococcus piscium). These results clearly demonstrate the importance of having an accurate bacterial named entity recognition method and sentence segmentation methods in the end-to-end bacterial interaction extraction pipeline.

\begin{table}[]
\label{tab:e2e}
\centering

\caption{The contribution of each of the components to the errors of the end-to-end pipeline. P.S. If a component is not there, the gold standard annotations are used for that component and only the other components of the pipeline are run. For e.g. in IE + NER, the gold standard segmented sentences are used, named entities are predicted by NER model and interactions are predicted by IE model. Similarly in Only IE, ground truth segmented as well as NER annotated sentences are used, and only the final interactions are predicted by the IE model.
}
\begin{tabular}{|l|l|l|l|}
\hline
              & Precision & Recall & F1   \\ \hline
Only IE       & 0.89      & 0.93   & 0.91 \\ \hline
IE + NER      & 0.63      & 0.84   & 0.72 \\ \hline
SS + IE + NER & 0.62      & 0.84   & 0.71 \\ \hline
\end{tabular}
\end{table}

\subsection{Case Study and Error Analysis}
The models and pipelines that were developed can be used for multiple purposes. One potential application area is to validate statistically predicted relations among microbes, obtained using metagenomic data analysis. This forms the basis of the case study and is detailed further. 

In the domain of microbiome research, one commonly used methodology is finding the associations or interactions among pairs of bacterial taxa using cross-sectional or longitudinal metagenomic data. Such statistical methods often predict erroneous or spurious associations \cite{compositional}. Therefore, there is a need to validate such reported associations using preexisting information present in biomedical literature. 

\subsubsection{Methodology for Case Study}
The initial starting point is an interaction network of microbes obtained from a cross sectional microbiome study of Americans' gut, which has been detailed in \cite{yooseph}. The network is composed of 155 pairwise associations among taxa. Our case study is composed of the following steps:

1. Using each of the predicted interacting pairs of bacterial taxa, we conduct a text search to identify a number of scientific literature sentences mentioning the two bacterial taxa. Litsense \cite{litsense} which aims at semantic search was used.

2. The end-to-end pipeline was run on the sentences obtained from the previous step. The interacting pairs of bacterial mentions which match our initial pair of bacterial taxa are captured, along with the sentences they occur in.

3. The above two steps are repeated for every pair of interacting/associated bacterial taxa. Using this compiled data, a list of sentences corresponding to every pair of predicted associations. This list of sentences is hereby referred to as \textit{probable sentences}.

Using the above pipeline, sentences reporting interactions among 29 bacterial pairs (out of the starting 155) can be obtained. These 29 associated pairs of taxa had at least one probable sentences reporting some relation. Therefore, it can be assumed that these pairs have a high probability of having an actual interaction.

\subsubsection{Error Analysis}

However, not all of the interacting pairs as suggested by the \textit{probable sentences} are true predictions. Manual analysis of the \textit{probable sentences} revealed that 79\% of the predictions were correct i.e. 21\% of the relations were false positives, i.e. the models predict an interaction between a pair of taxa, whereas, manual analysis of the sentence reveals the contrary. Despite manual analysis, no systematic errors, i.e. repeated error patterns were found. A few of the errors are tabulated in Table \ref{tab:error_examples}. Using the above analysis only an idea about the false positive rates or precision could be obtained. However, the false negative rate or recall of the methodology could not be analysed.

\begin{table*}
\label{tab:error_examples}
\caption{Example of errors made by the pipeline.}
\resizebox{\columnwidth}{!}{
\begin{tabular}{|l|l|l|l|}
\hline
  & Sentence                                                                                                                                                                                                                                                         & Entity 1      & Entity 2         \\ \hline
1 & \begin{tabular}[c]{@{}l@{}}Slight, but significant (p \textless 0.05) positive correlation was identified between Roseburia and \\ Eubacterium and ASMI, indicating that these genera are less abundant in individuals with \\ smaller muscle mass.\end{tabular} & Roseburia     & Eubacterium      \\ \hline
2 & \begin{tabular}[c]{@{}l@{}}In addition, significant correlation was found between Anaerotruncus, Intestinimonas and \\ Oscillibacter and steroid and terpenoid biosyntheses.\end{tabular}                                                                        & Oscillibacter & Anaerotruncus    \\ \hline
3 & \begin{tabular}[c]{@{}l@{}}Out of these families, Bacteroidaceae with genera Bacteroides or Mediterranea represent \\ one of the most frequent Gram-negative colonisers of the distal intestinal tract.\end{tabular}                                             & Bacteroides   & Mediterranea     \\ \hline
4 & \begin{tabular}[c]{@{}l@{}}In the present study, we observed co-clustering based on similar dynamics of Bacteroides \\ and Parabacteroides species with Fusicatenibacter saccharivorans, a species within the \\ family Lachnospiraceae.\end{tabular}            & Bacteroides   & Fusicatenibacter \\ \hline
\end{tabular}
}
\end{table*}

\section{Discussions}

Extraction of bacterial interactions reported in biomedical literature can help scientists better understand and exploit the microbiome for a variety of use cases. In this paper, a method for the extraction of pairwise interactions among bacteria is illustrated. These pairwise interactions can be used to construct or supplement knowledge graphs \cite{kg1} in the domain of microbiology. In this aspect it is important to note that the BactInt datasets and interaction extraction methods trained using it can only predict an interaction among bacterial taxa. However, it cannot predict the type of interaction, i.e. whether the interaction is positive or negative or something more complex. In order to predict the interactions along with the type of interaction, more detailed datasets need to be created, similar to \cite{r14}. In order to construct a robust knowledge graph, the predictions made by the model should ideally be further inspected by a manual or crowd-sourcing effort.

For most deep learning models, one of the major factors that determines their accuracy and generalization ability is the size and variety of the training datasets. While constructing the BactInt Sentence Corpus, emphasis was put on developing a large corpus capturing diverse types of bacterial interactions. Manual annotation of a large corpus is time consuming, and in the domain of biomedical NLP, very expensive, as it needs to be done by trained biomedical experts. Therefore, the majority of biological domain NLP tasks has much smaller datasets compared to their general domain counterparts. Transfer learning can partially help alleviate the problem of small sized training data. In this work, a few publicly available datasets were used for explicit transfer learning and the utility of the method was demonstrated in the results. All the chosen datasets involve extracting interactions among the same type of entities e.g. protein-protein interactions or gene-regulation (where one gene is regulated by another) or drug-drug interactions. This is relatively similar to the problem of bacterial interaction extraction, in which interactions among the same entity type (bacteria) need to be extracted. However, certain domain specific interaction patterns make bacteria-bacteria interactions unique in their own way necessitating the enrichment of transfer learning with domain based learning. Several datasets (annotated corpora) for problems like microbe-disease association extraction \cite{r11} and bacteria-biotope extraction \cite{r10}, which are specific to the domain of microbiology, are also available publicly. Including these datasets in the transfer learning training phase may improve the performance on the task of extracting bacterial interactions. However, a thorough evaluation is necessary for such a protocol.

One interesting observation was the better performance of the BioBERT models over the BioGPT ones in the interaction extraction task. However, the BioGPT model (and GPT models in general) are perform quite well on many NLP tasks including information extraction \cite{biogpt}. One reason for this might be that in this paper, the BioGPT model was used as a discriminative (classifier) model rather than a generative (end-to-end) model \cite{biogpt}. BioGPT is pretrained like a language model and whereas it performs best on generative NLP tasks, its performance on discriminative tasks such as classification may be limited as compared to BioBERT. One interesting future work would be using BioGPT as a generative end-to-end model for bacterial interaction extraction.

While analysing the errors of the end-to-end pipeline, one interesting observation was that the bacterial NER model often predicted all the scientific names of taxa present in the text as bacterial named entities. This exposes a systemic drawback in the BioBERT BNER model \cite{bacnerlm}, i.e. it does not really identify bacterial names (which is the intended task), but all types of scientific names. One alternative to trained machine learning models for bacterial NER is a simple dictionary lookup, using all bacterial names from NCBI Taxonomy \cite{ncbitaxonomy} or similar sources. However, this would lead to reduction in accuracy as several bacterial strain and substrain names are not present in the taxonomies, and dictionaries would be unable to take into account slight variations in bacterial names.

\section{Conclusion}

In this paper, a dataset for extracting bacterial interactions from biomedical scientific literature text was introduced. Additionally, a language model (BioBERT) based method to extract interactions from bacterial named entity marked sentences was described, along with a transfer learning based strategy to significantly improve the performance using other interaction extraction datasets. Finally, an end-to-end pipeline for the extraction of bacterial interactions from a given biomedical text (multiple sentences without marked bacterial named entities) was described and the performance was evaluated. The sources of errors in the pipeline were also analysed. We expect the dataset and the strategies described in this paper will lead to the further development of novel interesting strategies for the extraction of bacterial interactions from biomedical literature, and in turn be used to acquire and consolidate usable structured knowledge regarding bacterial interactions, thereby adding to the body of knowledge in the domain of microbiology. Interested readers are requested to get in touch with the corresponding author for access to the data.

\bibliographystyle{plain}

\bibliography{main}

\end{document}